\documentclass[preprint2]{aastex}
\usepackage{multirow}
\usepackage{float}
\usepackage{subfigure}

\slugcomment{Submitted to the Astrophysical Journal.}
\shorttitle{Hot Methane Line Lists}
\shortauthors{Hargreaves et al.}

\begin{document}

\title{Hot methane line lists for exoplanet and brown dwarf atmospheres\\ (CORRECTED\altaffilmark{\dag})}

\author{Robert J. Hargreaves\altaffilmark{1}}
\affil{Department of Chemistry, University of York, Heslington, York, YO10 5DD,
UK; \email{rjh135@york.ac.uk}}

\author{Christopher A. Beale\altaffilmark{1}}
\affil{Department of Physics, University of York, Heslington, York, YO10 5DD,
UK; \email{cbeale@odu.edu}}

\author{Laurent Michaux}
\affil{Department of Physics, University of York, Heslington, York, YO10 5DD,
UK; \email{lm595@york.ac.uk}}

\author{Melis Irfan\altaffilmark{2}}
\affil{Department of Physics, University of York, Heslington, York, YO10 5DD,
UK; \email{melis.irfan@postgrad.manchester.ac.uk}}

\and

\author{Peter F. Bernath}
\affil{Department of Chemistry \& Biochemistry, Old Dominion University, 4541 Hampton Boulevard, Norfolk, VA, 23529-0126, USA; and Department of Chemistry, University of York, Heslington, York, YO10 5DD,
UK; \email{pbernath@odu.edu}}

\altaffiltext{\dag}{ It was brought to our attention that the supplementary material of the original article (2012, \textit{ApJ}, 757, 46) contained a mistake. The correct data is presented here and is also available as an erratum (2013, \textit{ApJ}, 774, 89, doi:10.1088/0004-637X/774/1/89).}
\altaffiltext{1}{ currently at Department of Chemistry \& Biochemistry, Old Dominion University,
4541 Hampton Boulevard, Norfolk, VA, 23529-0126, USA.}
\altaffiltext{2}{ currently at School of Physics and Astronomy, The University of Manchester,
Oxford Road, Manchester, M13 9PL, UK.}

\begin{abstract}
We present comprehensive experimental line lists of methane (CH$_{4}$) at high temperatures obtained by recording Fourier transform infrared emission spectra. Calibrated line lists are presented for the temperatures 300 -- 1400$^{\circ}$C at twelve 100$^{\circ}$C intervals spanning the 960 -- 5000 cm$^{-1}$ (2.0 -- 10.4 $\mu$m) region of the infrared. This range encompasses the dyad, pentad and octad regions, i.e., all fundamental vibrational modes along with a number of combination, overtone and hot bands. Using our CH$_{4}$ spectra, we have estimated empirical lower state energies ($E_{low}$ in cm$^{-1}$) and our values have been incorporated into the line lists along with line positions ($\tilde{\nu}$ in cm$^{-1}$) and calibrated line intensities ($S'$ in cm molecule$^{-1}$). We expect our hot CH$_{4}$ line lists to find direct application in the modeling of planetary atmospheres and brown dwarfs.
\end{abstract}


\section{INTRODUCTION}

Methane (CH$_{4}$) is the most fundamental alkane molecule and is second only to CO$_{2}$ in causing climate change \citep{ipcc07}. The presence of CH$_{4}$ in the Earth's atmosphere is a consequence of both anthropogenic (e.g., landfills, rice cultivation, waste water treatment) and natural sources (e.g., methanogens, wetlands, wildfires). CH$_{4}$ can also be found throughout the solar system \citep{karkoschka94} including the atmospheres of Jupiter, Saturn, Uranus, Neptune and tentatively in the Martian atmosphere \citep{mumma09}. Titan, the largest moon of Saturn, has a dense atmosphere which is mostly made up of nitrogen but contains a relatively large proportion of CH$_{4}$ (approximately 1.4\% in the stratosphere; Niemann et al. 2005). The concentration of CH$_{4}$ is higher near the surface and it is believed a methane cycle exists which leads to the formation of liquid CH$_{4}$ lakes \citep{stofan07}. Interestingly, CH$_{4}$ has also been observed in the spectral energy distributions (SEDs) of cool substellar objects such as brown dwarfs \citep{cushing06} and more recently in exoplanet atmospheres \citep{swain08, seadem10}.

Cool stars and brown dwarfs are classified according to the molecular features in the observed SED \citep{bernath09}. The cooler the object, the more molecular signatures can be identified in the SED \citep{kirkpatrick05}. L dwarfs display near-infrared electronic transitions of metal hydrides such as FeH \citep{hargreaves10} and weaker TiO and VO bands \citep{geballe02} which are more commonly associated with M dwarfs \citep{leggett01,bernath09}. T dwarfs contain significant absorption due to H$_{2}$O and CH$_{4}$ \citep{cushing06} and have been referred to as `methane dwarfs' \citep{hausc09} because methane is the characteristic absorber. The newest category of brown dwarf (Y class) has recently been observed \citep{cushing11} and displays characteristic NH$_{3}$ absorptions along with additional CH$_{4}$ absorption. Given these band assignments, the SEDs of brown dwarfs are complex and still contain numerous unassigned features.

The study of exoplanets is relatively new, as it was not until the mid-1990s that the first exoplanet was discovered \citep{mandq95}. In recent years, `hot-Jupiters' are being discovered at an increasing rate and \citet{kirkpatrick11} were able to identify a further 100 using NASA's \textit{Wide-field Infrared Survey Explorer}. It is now possible to probe the atmospheres of exoplanets \citep{char02} by utilizing the natural transit as the planet passes in front of its parent star. The resulting flux undergoes a characteristic `dip' at specific wavelengths during the planetary transit \citep{char00} and from the extremely small variations in the wavelength dependence of the dips it is possible to determine which molecules are present. By utilizing the transit method, H$_{2}$O \citep{barman08, grillmair08}, CH$_{4}$ \citep{swain08}, CO \citep{swain09b} and CO$_{2}$ \citep{swain09a, tinetti10} have already been identified. Currently this method has only been successful for large hot-Jupiters, but NASA's \textit{Kepler} mission is discovering ever smaller planets which more closely resemble Earth, such as \textit{Kepler-22b} \citep{borucki12}. Exoplanets are now thought to be the rule, rather than the exception, throughout the Milky Way \citep{cassan12} and the ultimate exoplanet goal is to detect the distinctive signatures of extra terrestrial life through observation of the atmospheres of these distant planets.

\begin{deluxetable}{lccl}
\tabletypesize{\small}
\tablewidth{0pt}
\tablecaption{The fundamental infrared modes of CH$_{4}$\label{tab:modes}}
\tablehead{\colhead{Mode} & \colhead{Degeneracy} & \colhead{Band Origin\tablenotemark{a} (cm$^{-1}$)} & \colhead{Type}}
\startdata
$\nu_{1}$ $(a_{1})$  & 1  & 2916.481 & Symmetric C--H Stretch     \\
$\nu_{2}$ $(e)$      & 2  & 1533.333 & Bend                       \\
$\nu_{3}$ $(t_{2})$  & 3  & 3019.493 & Antisymmetric C--H Stretch \\
$\nu_{4}$ $(t_{2})$  & 3  & 1310.761 & Bend                       \\
\enddata
\tablenotetext{a}{Taken from \citet{albert09}.}
\end{deluxetable}

Infrared spectroscopy of CH$_{4}$ is complex due to the high symmetry of the molecule and the polyad structure which arises from close-lying vibrational levels \citep{boudon06}. CH$_{4}$ is a tetrahedral molecule (T$_{d}$ symmetry) and is a classic example of a spherical top. There are $3N-6=9$ infrared modes with four stretches and five bends. The four C--H stretches correspond to a symmetric $a_{1}$ stretch ($\nu_{1}$ at 2916.481 cm$^{-1}$) and a triply degenerate $t_{2}$ antisymmetric stretch ($\nu_{3}$ at 3019.493 cm$^{-1}$). The five bends correspond to a doubly degenerate $e$ bending mode ($\nu_{2}$ at 1533.333 cm$^{-1}$) and a triply degenerate $t_{2}$ bending mode ($\nu_{4}$ at 1310.761 cm$^{-1}$). The spectroscopic constants of CH$_{4}$ are summarised in Table~\ref{tab:modes} \citep{albert09, herzberg91} and only the $\nu_{3}$ and $\nu_{4}$ fundamental modes are nominally infrared active; however, it is possible for other modes to gain intensity through strong interactions within the polyad structure. The interesting polyad structure of the infrared spectrum of CH$_{4}$ arises because the bending frequencies are approximately equal ($\nu_{2}\approx\nu_{4}$), the stretching frequencies are also approximately equal ($\nu_{1}\approx\nu_{3}$) and approximately twice the bending frequencies \citep{albert09}. For example, the pentad region near 3 $\mu$m has 5 vibrational bands since $\nu_{1}\approx\nu_{3}\approx2\nu_{2}\approx2\nu_{4}\approx\nu_{2}+\nu_{4}$.

CH$_{4}$ has also frequently been the target of high quality theoretical calculations based on the solution of the vibration-rotation Schr\"{o}dinger equation \citep{wands02,wandc04,cassamchenai03} using \textit{ab initio} potential energy surfaces \citep{sandp01,schwenke02,nikitin11}. The accuracy of vibration-rotational calculations are typically far from experiment; however, recent calculations of nominally forbidden pure rotational lines \citep{cassamchenai12} are able to approach experimental accuracy.

\citet{albert09} have reported a study of CH$_{4}$ in the 0 -- 4800 cm$^{-1}$ region and these lines form the basis of the HITRAN 2008 database \citep{rothman09}. Extensive work has been carried out in Grenoble including the preparation of a cold CH$_{4}$ line list \citep{mondelain11, wang11, wang12, campargue12a} which has been successfully applied to the atmosphere of Titan \citep{campargue12b}. Recently focus has shifted to the generation of hot line lists for use in brown dwarfs and exoplanet models, either from \textit{ab initio} calculations \citep{warm09} or from experiment \citep{nassar03,perrin07,thievin08}.

Radiative transfer models are typically used to identify molecules in the atmospheres of exoplanets \citep{swain08}. Adaptations for hot-Jupiter atmospheres include known strongly absorbing molecules (i.e. H$_{2}$O and CH$_{4}$) and one of the options for CH$_{4}$ has been the HITRAN 2008 database \citep{rothman09} combined with experimental data obtained by \citet{nassar03}. HITRAN covers a broad spectral range but is intended for room temperature applications whereas the \citet{nassar03} data are only recorded at three higher temperatures (800, 1000 and 1273 K). This limits the accuracy of the modeled SEDs and leads to errors in deriving effective temperatures \citep{hausc09}. We hope to address this issue by providing a series of high-resolution, hot CH$_{4}$ line lists at temperatures relevant to exoplanet and brown dwarf atmospheres with the inclusion of empirical lower state energies where possible. This work will cover the dyad ($\nu_{2}$, $\nu_{4}$), pentad ($\nu_{1}$, $\nu_{3}$, $2\nu_{2}$, $2\nu_{4}$, $\nu_{2}+\nu_{4}$) and octad ($3\nu_{2}$, $3\nu_{4}$, $\nu_{1}+\nu_{2}$, $\nu_{1}+\nu_{4}$, $\nu_{3}+\nu_{2}$, $\nu_{3}+\nu_{4}$, $2\nu_{2}+\nu_{4}$, $\nu_{2}+2\nu_{4}$) regions of the infrared including numerous hot bands. Similar work has already been carried out on NH$_{3}$ \citep{hargreaves11, hargreaves12}. Our data sets, and molecular opacities derived from them, can be used directly in exoplanet and brown dwarf models.

\section{EXPERIMENTAL METHOD}

\begin{figure*}
\centering
\includegraphics[angle=90,scale=0.69]{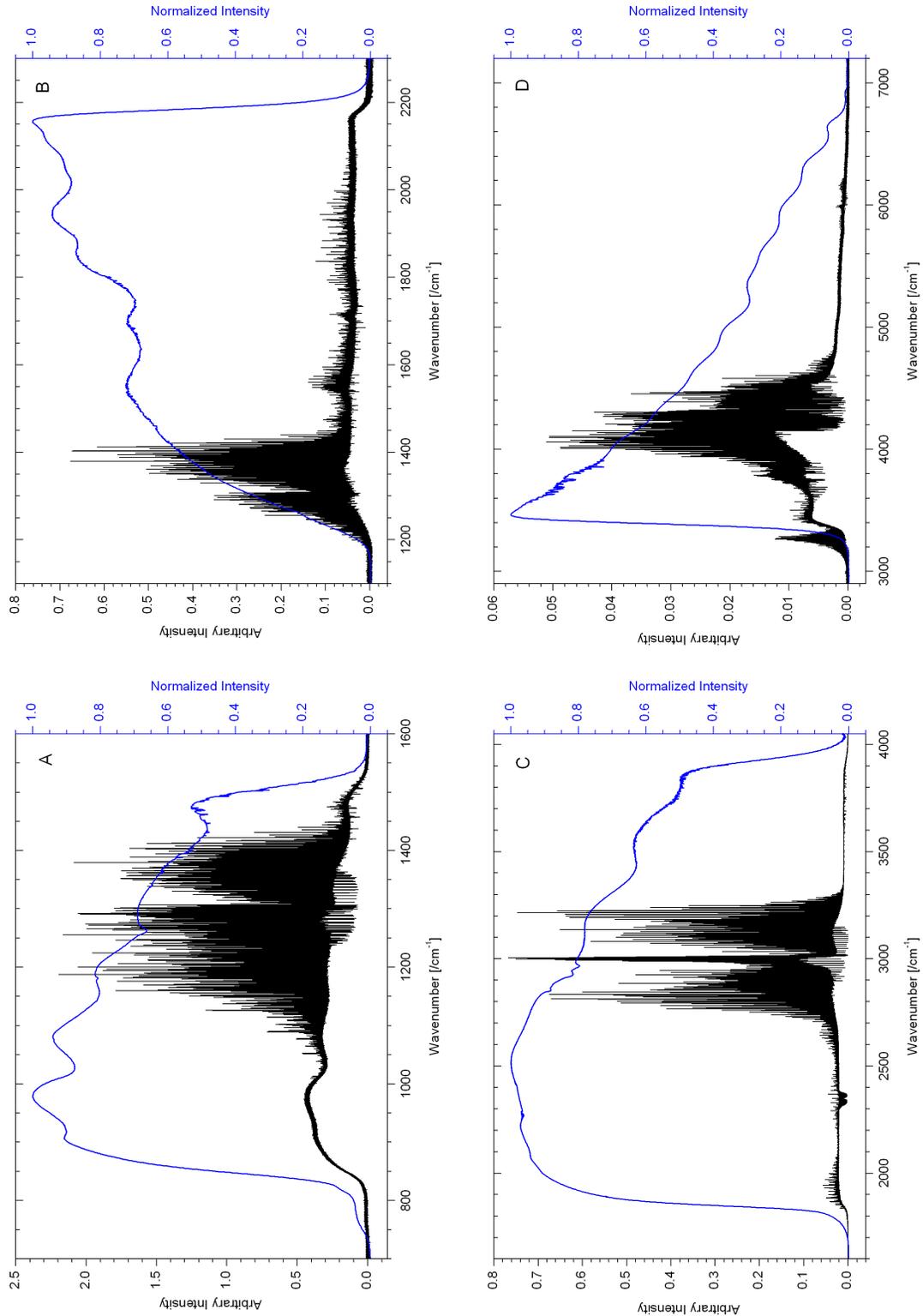}
\caption{High-resolution laboratory emission spectra of CH$_{4}$ at 1000$^{\circ}$C for all four regions (black) with the corresponding normalized instrument response function (blue). The baseline is caused by residual thermal emission from the tube walls and a small amount of absorption due to cold CH$_{4}$, CO$_{2}$ ($\nu_{3}$ near 2350 cm$^{-1}$) and H$_{2}$O can be seen.\label{fig:allspectra}}
\end{figure*}

High-resolution laboratory emission spectra of hot CH$_{4}$ were recorded in four separate parts in order to maximise the signal-to-noise for each spectral region. The four spectral regions are shown in Figure~\ref{fig:allspectra} for 1000$^{\circ}$C. A similar setup was used in our previous work on NH$_{3}$ \citep{hargreaves11, hargreaves12} and only key experimental features will be summarized here.

The experiment coupled a Fourier transform infrared (FT-IR) spectrometer with a temperature controllable tube furnace capable of maintaining stable temperatures up to 1400$^{\circ}$C with an accuracy $\pm10^{\circ}$C. The tube furnace surrounds the central portion of an alumina (Al$_{2}$O$_{3}$) tube which is sealed at both ends by infrared windows. CH$_{4}$ gas is allowed to flow through the alumina tube at a stable pressure in order to avoid the build up of impurities. The infrared emission from the alumina tube was focussed onto the aperture of the FT-IR spectrometer using a lens. The small region between the spectrometer and tube furnace was purged with dry air in order to minimize H$_{2}$O absorption lines in the recorded spectra. Table~\ref{tab:conditions} contains the experimental parameters used for all four regions.

\begin{deluxetable}{lcccc}
\tabletypesize{\small}
\tablewidth{0pt}
\tablecaption{Experimental conditions\label{tab:conditions}}
\tablehead{ & \colhead{Region A}  & \colhead{Region B}   & \colhead{Region C}   & \colhead{Region D}}
\startdata
Total file coverage (cm$^{-1}$)                &    600--1900  &   1000--3000 &   1500--5000  &   2500--8000\\
Temperature Range ($^{\circ}$C)                &    300--1400  &    300--1400 &    300--1400  &    500--1300\\
Detector                                       &          MCT  &          MCT &         InSb  &         InSb\\
Beam Splitter                                  &          KBr  &    CaF$_{2}$ &    CaF$_{2}$  &    CaF$_{2}$\\
Windows                                        &        KRS-5  &    CaF$_{2}$ &    CaF$_{2}$  &    CaF$_{2}$\\
Lens                                           &         ZnSe  &    CaF$_{2}$ &    CaF$_{2}$  &    CaF$_{2}$\\
Total scans                                    &          240  &          300 &          300  &          800\\
Resolution (cm$^{-1}$)                         &         0.01  &        0.015 &        0.015  &         0.02\\
Aperture (mm)                                  &          2.5  &          2.5 &          2.5  &          1.7\\
CH$_{4}$ pressure (Torr)\tablenotemark{a}      &          4.0  &          4.0 &          4.0  &         60.0\\
Zero-fill factor                                &   $\times16$  &   $\times16$ &   $\times16$  &   $\times16$\\
\enddata
\tablenotetext{a}{ This is an average pressure to the nearest 0.1 Torr.}
\end{deluxetable}

Spectra were recorded at 100$^{\circ}$C intervals between 300 and 1400$^{\circ}$C for regions A, B and C and between 500 and 1300$^{\circ}$C for region D as this was the temperature coverage of observable CH$_{4}$ emission. Each region was limited using an appropriate filter chosen to allow for overlap between neighbouring regions. The resolution of each spectral region was based on the Doppler and pressure broadening widths, and was selected to minimize the acquisition time whilst maintaining the maximum line resolution.

The temperature of the tube furnace was measured and maintained at high temperatures using a thermocouple combined with a programmable controller. The ends of the alumina tube required cooling to avoid damaging the rubber o-rings (which maintain the pressure seal) and the consequence of this is a temperature gradient within the tube. The observed infrared emission spectra are dominated by the high temperature central portion of the tube (see Figure~\ref{fig:allspectra}) and the emission lines are assumed to be formed at the same temperature as the furnace.

The emission lines in each spectrum were picked using the program WSpectra \citep{carleer01} and the line intensity of each line was obtained by fitting a Voigt line profile. The number of CH$_{4}$ emission lines recorded for each region is given in Table~\ref{tab:observed}.

\begin{deluxetable}{lccccc}
\tabletypesize{\small}
\tablewidth{0pt}
\tablecaption{Observed emission lines of CH$_{4}$\label{tab:observed}}
\tablehead{\colhead{Temperature ($^{\circ}$C)} & \colhead{Region A} & \colhead{Region B} & \colhead{Region C} & \colhead{Region D} & \colhead{Total}}
\startdata
300  &  4,365 & 2,589                      &  4,118                      &      -                     & 11,072 \\
400  &  6,656 & 3,566                      &  7,752                      &      -                     & 17,974 \\
500  &  7,950 & 4,725                      &  9,438                      &  5,242                     & 27,355 \\
600  &  9,524 & 6,021                      & 14,109                      &  7,766                     & 37,420 \\
700  & 10,381 & 6,344                      & 15,376                      & 10,067                     & 42,168 \\
800  & 11,189 & 6,951                      & 19,297                      & 12,696                     & 50,133 \\
900  & 11,607 & 7,276                      & 21,148                      & 14,688                     & 54,719 \\
1000 & 12,005 & 7,889                      & 21,042                      & 16,563                     & 57,499 \\
1100 & 12,105 & 7,667                      & 21,850                      & 16,062                     & 57,684 \\
1200 & 12,368 & 7,637                      & 22,511                      & 10,269\tablenotemark{a}    & 52,785 \\
1300 & 12,235 & 8,118                      & 23,332                      &  9,640\tablenotemark{a}    & 53,325 \\
1400 & 12,510 & 6,095\tablenotemark{b}     & 10,590\tablenotemark{b}     &      -                     & 29,195 \\
\enddata
\tablenotetext{a}{These spectra were contaminated by C$_{2}$H$_{2}$ and C$_{2}$H$_{4}$ emission lines and have not been included in our analysis as the impurity lines could not be completely removed.}
\tablenotetext{b}{These spectra were unusually noisy and as a result the number of lines observed is reduced.}
\end{deluxetable}

A system response function was measured for each region by recording a blackbody spectrum emitted from a graphite rod placed at the centre of the the alumina tube (i.e., at the centre of the furnace). This accounts for the relative contribution of the system (e.g., windows, lens, filters, etc.) to the intensity of the observed lines. The spectrum is compared to a theoretical blackbody at the temperature of the furnace and then normalized. These instrument response functions are shown in Figure~\ref{fig:allspectra} along with the CH$_{4}$ emission lines at 1000$^{\circ}$C.

It was necessary to calibrate both the wavenumber scale ($x$-axis) and the intensity scale ($y$-axis) using the HITRAN line list. The wavenumber scales were calibrated by selecting 20 strong, clear, symmetric lines and averaging the correction factor. Each temperature spectrum had similar calibration factors and typical values (for the 1000$^{\circ}$C spectra) were 1.000001351 for region A, 1.000001677 for region B, 1.000001051 for region C and 1.000000091 for region D. This results in a typical shift of 0.005 cm$^{-1}$ at 3000 cm$^{-1}$ for the $\nu_{3}$ band. We therefore expect the accuracy of each each line list to be $\pm$0.005 cm$^{-1}$ or better.

The intensity scales have also been calibrated according to HITRAN. First, each emission line list was converted into an arbitrary absorption scale to make them comparable to the intensities of the lines contained in HITRAN. This method has previously been used by \citet{nassar03} and more recently by \citet{hargreaves11, hargreaves12} by using the equation
\begin{equation}
S_{a} = \frac{ S_{e} }{ \nu^{3} \exp\left( -\frac {h \nu} {k T} \right)},
\end{equation}
where $S_{a}$/$S_{e}$ is the intensity of the absorbed/emitted line, $\nu$ is the frequency and $T$ is the temperature.

Each line list was then put onto the same intensity scale as recorded by region A, and comparisons to HITRAN were made. The HITRAN lines had to be scaled to the appropriate temperatures using
\begin{equation}
S'= S'_{0} \frac{Q_{0}}{Q} \exp\left(\frac {E_{low}} {k T_{0}} - \frac {E_{low}} {k T}\right) \left[\frac{1 - \exp\left(-\frac {h \nu_{10}} {k T}\right)}{1 - \exp\left(-\frac {h \nu_{10}} {k T_{0}}\right)}\right],
\label{eqn:ratios}
\end{equation}
where $S'$ is the intensity, $Q$ is the partition function, $T$ is the temperature, $\nu_{10}$ is the line frequency and  $E_{low}$ is the empirical lower state energy \citep{bernath05}. A zero subscript refers to the same parameters but at the reference temperature. The partition functions can be found in Table~\ref{tab:partition} and were calculated for each temperature using the TIPS$\_$2011.for program \citep{laraia11} based on calculations made by \citet{wenger08}.

\begin{deluxetable}{cc}
\tabletypesize{\small}
\tablewidth{0pt}
\tablecaption{Methane partition function\label{tab:partition}}
\tablehead{\colhead{Temperature (K)} & \colhead{Partition Function\tablenotemark{a}} }
\startdata
 296 &     590.52 \\
 573 &   1,856.19 \\
 673 &   2,648.82 \\
 773 &   3,745.99 \\
 873 &   5,267.13 \\
 973 &   7,372.47 \\
1073 &  10,276.30 \\
1173 &  14,266.42 \\
1273 &  19,723.43 \\
1373 &  27,161.91 \\
1473 &  37,273.63 \\
1573 &  50,996.68 \\
1673 &  69,611.60 \\
\enddata
\tablenotetext{a}{Calculated using the TIPS$\_$2011.for program \citep{laraia11} based on calculations made by \citet{wenger08}.}
\end{deluxetable}

Wavenumber matches between our data and the temperature-scaled HITRAN lines were made ($\pm$0.005 cm$^{-1}$) and these intensities were compared. A calibrating factor could then be deduced but as the frequency increased, the intensity calibration factors became larger and more inconsistent. Regions A, B, and C could all be matched without difficulty since these regions cover strong emission bands; however, region D proved to be a problem. In the end, it was decided that an intensity calibration function could be extrapolated into region D based on the matches from regions A, B and C. The intensity calibration used to convert the observed arbitrary intensity ($S'_{arb}$) to intensities in HITRAN units ($S'_{H}$) is
\begin{equation}
S'_{H} = [a(\tilde{\nu})^{2} + b\tilde{\nu} + c]S'_{arb},
\label{eqn:3}
\end{equation}
where $\tilde{\nu}$ is in cm$^{-1}$ and $S'_{H}$ is in cm molecule$^{-1}$. The calibration function is temperature dependent and was applied to all regions, the constants are given in Table~\ref{tab:calibrations}.

\begin{deluxetable}{cccc}
\tabletypesize{\small}
\tablewidth{0pt}
\tablecaption{Wavenumber calibration functions\label{tab:calibrations}}
\tablehead{\colhead{Temperature} & \colhead{$a$} & \colhead{$b$} & \colhead{$c$} \\ \colhead{($^{\circ}$C)} & \colhead{(/$10^{12}$ cm$^{3}$ molecule$^{-1}$)} & \colhead{(/$10^{15}$ cm$^{2}$ molecule$^{-1}$)} & \colhead{(/$10^{18}$ cm molecule$^{-1}$)} }
\startdata
 300 &  2.92 &  -9.75 & 12.00 \\
 400 &  1.48 &  -2.15 &  4.98 \\
 500 &  2.82 &  -6.67 &  9.03 \\
 600 &  3.62 &  -9.87 & 12.30 \\
 700 &  4.04 & -10.70 & 13.30 \\
 800 &  2.56 &  -3.35 &  6.02 \\
 900 &  3.24 &  -6.17 &  8.63 \\
1000 &  4.87 & -12.40 & 14.20 \\
1100 &  4.00 &  -7.77 &  9.33 \\
1200 &  4.07 &  -7.18 &  7.95 \\
1300 &  5.06 & -12.00 & 12.60 \\
1400 &  4.68 &  -7.48 &  7.29 \\
\enddata
\end{deluxetable}

Each region had to be combined to form a complete line list at each temperature. The optical filters used during the acquisition of each spectrum allowed the regions to overlap; the common lines in the overlapping regions have been used to bring all regions onto the same scale. The number of emission lines in the final line lists were maximized by appropriately choosing to combine the regions at 1500, 1830 and 3350 cm$^{-1}$. Because emission lines were only recorded above 960 cm$^{-1}$ and region D was intentionally truncated at 5000 cm$^{-1}$, the spectral regions are 960 -- 1500 cm$^{-1}$ (region A), 1500 -- 1830 cm$^{-1}$ (region B), 1830 -- 3350 cm$^{-1}$ (region C) and 3350 -- 5000 cm$^{-1}$ (region D), i.e., 960 -- 5000 cm$^{-1}$ in total.

\begin{figure}[t]
\epsscale{1}
\plotone{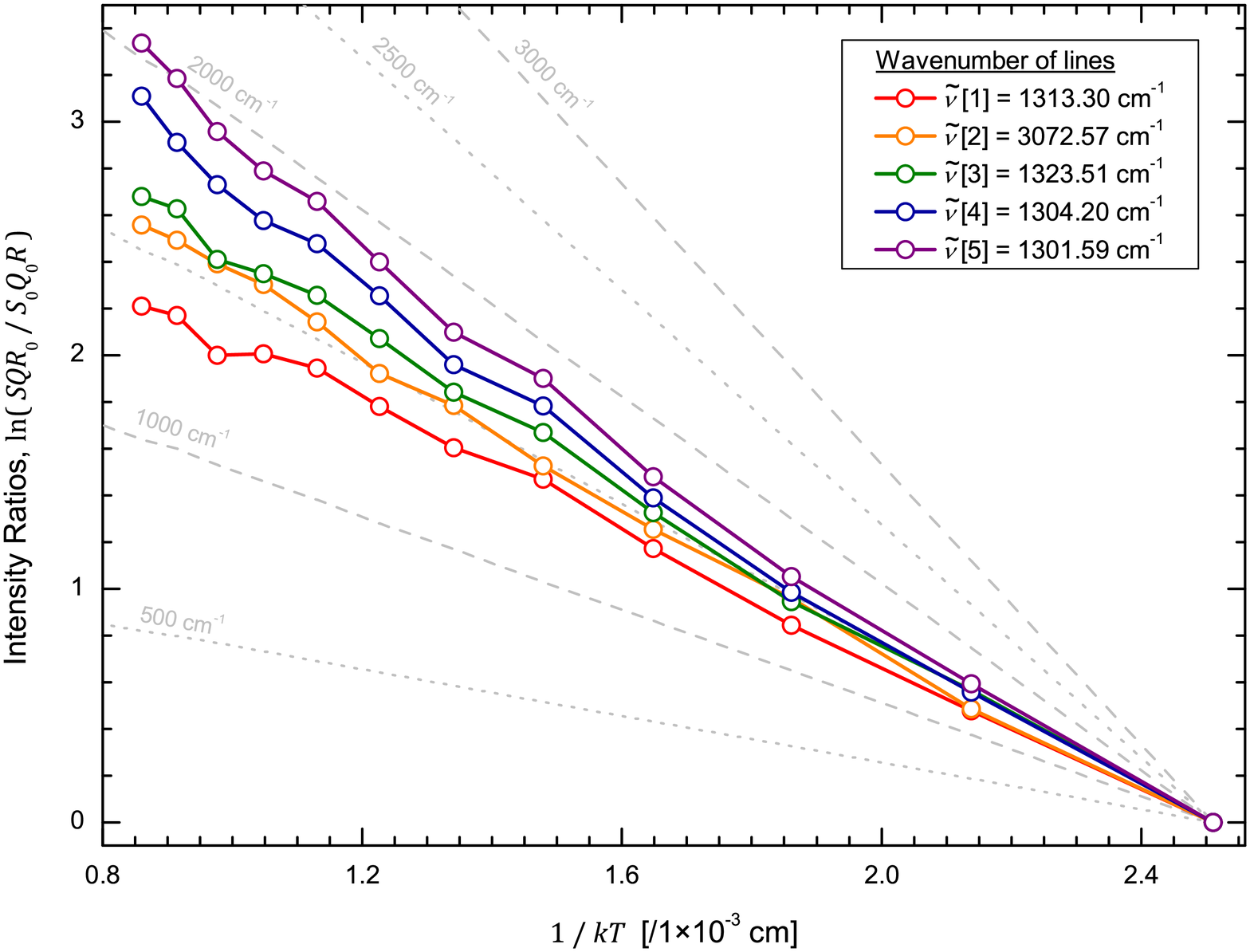}
\caption{Natural logarithm of the intensity ratios of five lines present in all 12 spectra are plotted against $1/kT$ (in cm), where $T$ is the temperature. The slope of each line directly yields the empirical lower state energy (in cm$^{-1}$) and for the lines plotted: $E_{low}[1]=1364.00$ cm$^{-1}$ (red), $E_{low}[2]=1590.79$ cm$^{-1}$ (orange), $E_{low}[3]=1644.41$ cm$^{-1}$ (green), $E_{low}[4]=1879.95$ cm$^{-1}$ (blue), and $E_{low}[5]=2036.83$ cm$^{-1}$ (purple).\label{fig:slope}}
\end{figure}

The emission lines for each temperature were compared to all other temperatures and all matchable lines were identified ($\pm$0.005 cm$^{-1}$). It is then possible to calculate the empirical lower state energy ($E_{low}$) of each line by rearranging Equation~\ref{eqn:ratios} so that
\begin{equation}
\ln\left(\frac{SQR_{0}}{S_{0}Q_{0}R}\right) = \frac {E_{low}} {k T_{0}} - \frac {E_{low}} {k T},
\label{eqn:4}
\end{equation}
where
\begin{equation}
R_{0} = {1 - \exp\left(-\frac {h \nu_{10}} {k T_{0}}\right)}.
\end{equation}
The slope of points plotted using Equation~(\ref{eqn:4}) for each line yields the $E_{low}$ and an example of this calculation is shown in Figure~\ref{fig:slope} for five CH$_{4}$ emission lines present at all temperatures.

This then forms the basis of the temperature dependant line lists which contain calibrated line positions ($\tilde{\nu}$ in cm$^{-1}$), calibrated line intensities ($S'$ in cm/molecule) and empirical lower state energies ($E_{low}$ in cm$^{-1}$). It was necessary to incorporate missing HITRAN lines into these line lists as explained in the results.

\section{RESULTS}

\begin{figure*}
\centering
\includegraphics[angle=90,scale=0.35]{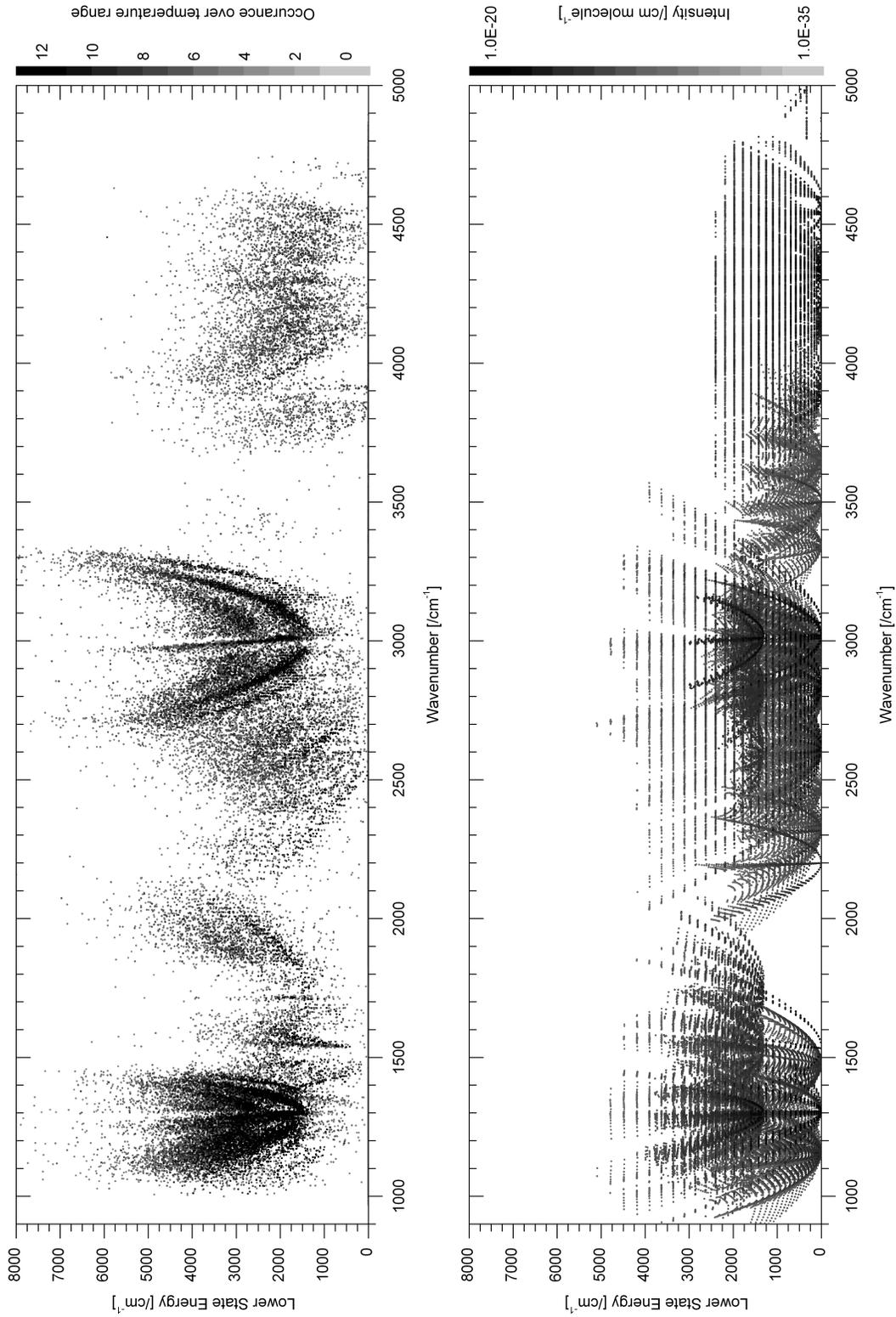}
\caption{Comparison of the empirical lower state energies (top panel) with the HITRAN lower state energies (bottom panel). The dyad ($\nu_{4}$ region), pentad ($\nu_{3}$ region) and octad ($\nu_{3}+\nu_{4}$) regions are clearly visible.\label{fig:elows}}
\end{figure*}

Using our method to calculate the $E_{low}$ values, it is possible to identify the dyad, pentad and octad infrared regions of CH$_{4}$ in the plots of $E_{low}$ versus wavenumber (Figure~\ref{fig:elows}).

The two most striking features are the $\nu_{4}$ region (dyad) around 1300 cm$^{-1}$ and the $\nu_{3}$ region (pentad) around 3020 cm$^{-1}$. The $\nu_{4}$ region contains lines of the fundamental $\nu_{4}$ bend and distinct $2\nu_{4}-\nu_{4}$ and $3\nu_{4}-2\nu_{4}$ hot bands at a $E_{low}$ of 1300 and 2600 cm$^{-1}$ respectively. The $\nu_{3}$ region contains a strong $P$-, $Q$-, and $R$-branch of the $\nu_{3}+\nu_{4}-\nu_{4}$ hot band above a $E_{low}$ of 1300 cm$^{-1}$ and parts of the fundamental $P$-, and $R$-branch of the $\nu_{3}$ mode. Another hot band, possibly the $2\nu_{3}-\nu_{3}$, can be seen above a $E_{low}$ of $\sim$3000 cm$^{-1}$. Less obvious but still present are tentative observations of the $Q$-branch of the `forbidden' $\nu_{2}$ fundamental bend at approximately 1530 cm$^{-1}$. Also present are features of the $\nu_{3}+\nu_{4}$ combination region (octad) around 4300 cm$^{-1}$, although the features in this region are less distinct as it is congested by many other bands. Comparisons to HITRAN for this region shows a complex band structure as expected due to the high density of transitions.

There is reasonable agreement between the empirical $E_{low}$ values and the HITRAN $E_{low}$ values (Figure~\ref{fig:elows}) within the $\nu_{3}$ and $\nu_{4}$ region. A hot band at a lower state energy of approximately 3000 cm$^{-1}$ in the pentad region (tentatively assigned the $2\nu_{3}-\nu_{3}$ hot band) is absent in HITRAN and is not very well defined in our data. We have observed emission lines with higher rotational levels than are present in HITRAN which is most apparent in the vicinity of the $\nu_{3}$ region. The HITRAN database clearly contains many CH$_{4}$ lines, but the majority of $E_{low}$ values have been obtained from calculations; these $E_{low}$ values are visible in Figure~\ref{fig:elows} as horizontal `stripes' which correspond to the distinct rotational levels.

\begin{figure*}
\centering
\subfigure[Cluster splittings and $E_{low}$ values from HITRAN]{%
\label{fig:cluster:A}
\includegraphics[width=0.85\textwidth]{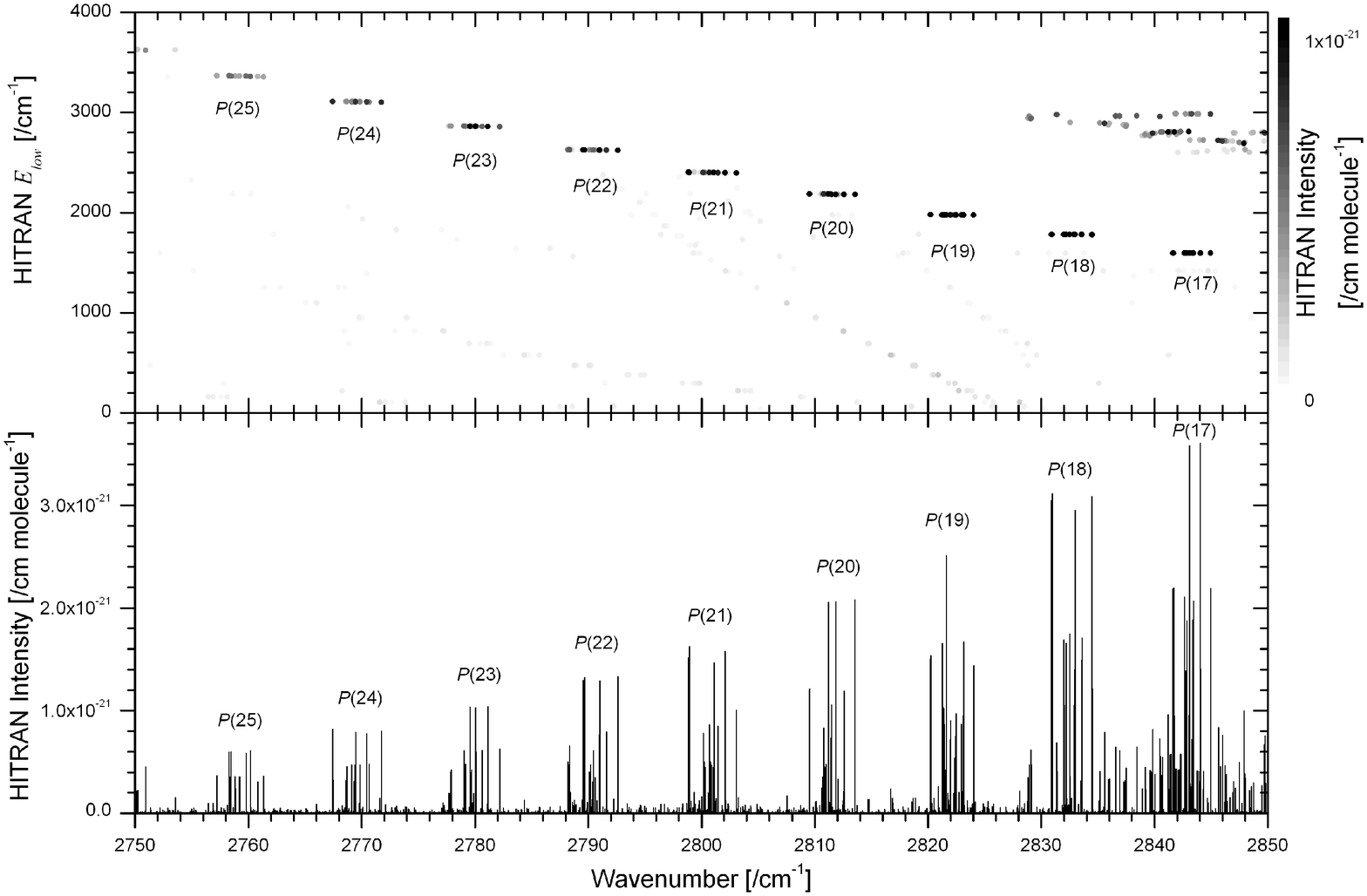}
}\\%
\subfigure[Cluster splittings and empirical $E_{low}$ values]{%
\label{fig:cluster:B}
\includegraphics[width=0.85\textwidth]{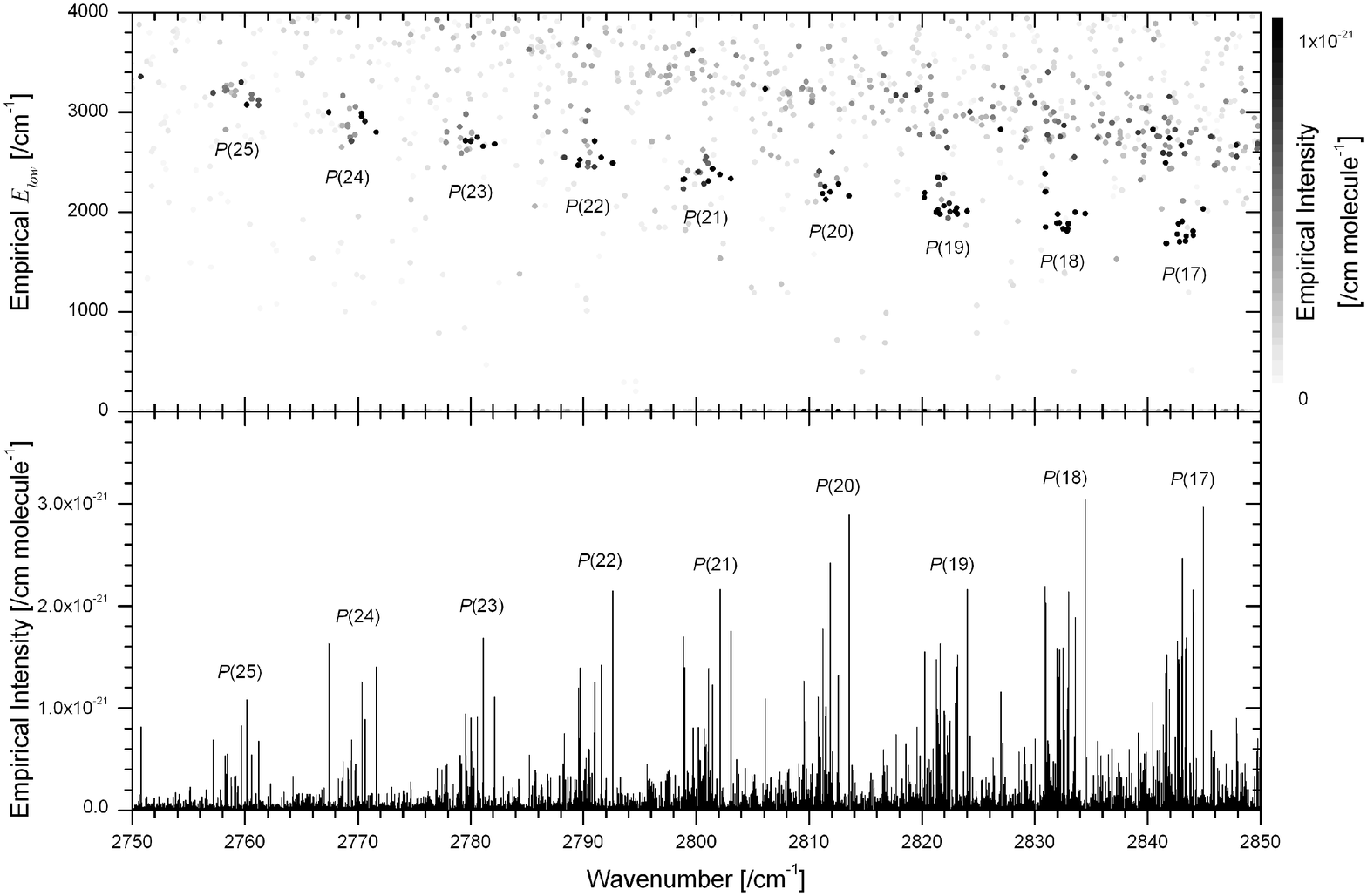}
}%
\caption{Cluster splitting of the $P$-branch lines of the $\nu_{3}$ mode of CH$_{4}$. Panel (a) contains the HITRAN $E_{low}$ values plotted with the scaled HITRAN lines at 1000$^{\circ}$C and panel (b) contains the empirical $E_{low}$ values plotted with the calibrated lines at 1000$^{\circ}$C. The grey scale of the $E_{low}$ value is taken from the intensity of the line at 1000$^{\circ}$C.\label{fig:cluster:both}}
\end{figure*}

Our method is also able to distinguish rotational structure of the branches of the $\nu_{3}$ fundamental mode as shown in Figure~\ref{fig:cluster:both}. The rotational levels of a rigid spherical top are given by $BJ(J+1)$ but there is a $(2J+1)^{2}$-fold degeneracy for each $J$ (i.e., $(2J+1)$ from the $M$ quantum number and $(2J+1)$ from $K$). As the molecule rotates and distorts, the $K$ degeneracy is partly lifted and the lines show a `cluster-splitting'. Figure~\ref{fig:cluster:both} highlights the observed structure which can be seen in the calculated $E_{low}$ values of the $\nu_{3}$ region of CH$_{4}$.

It is worth noting that points below an $E_{low}$ of approximately 1000 cm$^{-1}$ are absent from Figure~\ref{fig:elows} due to the temperature gradient within the alumina tube. The CH$_{4}$ in the cooler ends of the tube absorbs the low-$J$ emission lines of the fundamental bands and accounts for the fact that only high-$J$ levels are seen in the branches of the $\nu_{3}$ mode as shown in Figures~\ref{fig:elows} and~\ref{fig:cluster:both}.

Self-absorption appears in our emission spectra as either complete absorption, resulting in a missing line, or as partial absorption which gives a distinctive double peak \citep{hargreaves11}. In order to provide more complete line lists we have added the missing lines from HITRAN to our line lists\footnote{This was the source of the original error. The correct data is presented here and is also available as an erratum (2013, \textit{ApJ}, 774, 89, doi:10.1088/0004-637X/774/1/89).}. We therefore incorporate HITRAN lines ($\pm$0.0025 cm$^{-1}$) with an intensity greater than that of a threshold value at each temperature. Our sensitivity decreases toward higher wavenumbers (regions A, C and D from Figure~\ref{fig:allspectra}) and as a result the cut-off threshold is not a constant and changes over each region in order to match the sensitivity of the observations. The effect of adding HITRAN lines to our line lists is shown in Table~\ref{tab:breakdown} which lists the added HITRAN lines and the observed emission lines. In order to avoid double counting of lines, all lines in the emission line list within $\pm$0.0025 cm$^{-1}$ of a HITRAN line (with an intensity above the selection threshold) were deleted and replaced by the HITRAN lines.

\begin{deluxetable}{cccc}
\tablenum{6}
\tabletypesize{\small}
\tablewidth{0pt}
\tablecaption{Line list summary\label{tab:breakdown}}
\tablehead{\colhead{Temperature ($^{\circ}$C)} & \colhead{Observed Lines} & \colhead{Added HITRAN Lines} & \colhead{Total Lines} }
\startdata
 300 &   3,255  &   9,120  &  12,375  \\
 400 &   6,542  &  14,803  &  21,345  \\
 500 &  12,437  &  21,518  &  33,955  \\
 600 &  19,247  &  29,471  &  48,718  \\
 700 &  23,644  &  31,249  &  54,893  \\
 800 &  30,368  &  37,708  &  68,076  \\
 900 &  34,929  &  38,218  &  73,147  \\
1000 &  37,706  &  38,559  &  76,265  \\
1100 &  39,485  &  39,613  &  79,098  \\
1200 &  28,532  &  30,419  &  58,951  \\
1300 &  30,189  &  28,218  &  58,407  \\
1400 &  19,788  &  17,067  &  36,855  \\
\enddata
\end{deluxetable}

\begin{figure}[t]
\epsscale{1}
\plotone{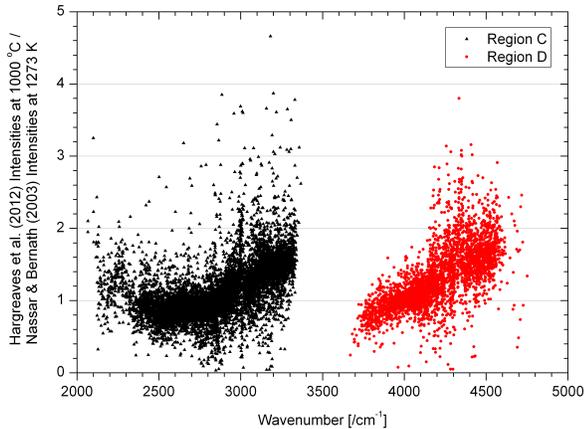}
\caption{Comparison of line intensities at 1000$^{\circ}$C with the values for corresponding lines (within $\pm$0.0025 cm$^{-1}$) recorded by \citet{nassar03} at 1273 K.\label{fig:nassar}}
\end{figure}

Comparisons made with \citet{nassar03} at 1000$^{\circ}$C ($\pm$0.0025 cm$^{-1}$) suggest our intensities are accurate to within a factor of two (Figure~\ref{fig:nassar}). To estimate the error in the empirical $E_{low}$ values they were compared to those found in HITRAN for a number of lines which occur in both data sets. Table~\ref{tab:errors} shows that the percentage difference between the empirical $E_{low}$ calculation and HITRAN for the lines plotted in Figure~\ref{fig:slope} is less than 10 \%. The examples shown are lines which appear in all spectra and as a result are the most reliable. These particular lines have also been removed from our final line lists because they coincide with HITRAN lines. As the number of points used to calculate the lower state energy decreases, the error increases. It is therefore difficult to quote an overall error for our $E_{low}$ values  but we aim to represent the error by the inclusion of a quality factor within the line lists. `H' signifies that the line position, intensity and lower state energy has been inserted from the HITRAN database. `1' means that the lower state energy has been calculated from the intensities of a line that appears in 10 or more spectra. `2' refers to any line which appears in 6 to 9 spectra and `3' denotes that the lower state energy has been determined from 3 to 5 spectra. Finally, a `0' indicates that the lower state energy could not be determined from the spectra available; in this case it means that the line occurs in less than 3 spectra in total. It should be noted that for region D a maximum quality factor of only `2' is achievable. An extract of the hot CH$_{4}$ line list at 1000$^{\circ}$C is displayed in Table~\ref{tab:extract}, the complete line lists are available from the website of the Astrophysical Journal\footnote{(2013, \textit{ApJ}, 774, 89, doi:10.1088/0004-637X/774/1/89)}.

\begin{deluxetable}{ccccc}
\tablenum{7}
\tabletypesize{\small}
\tablewidth{0pt}
\tablecaption{Empirical lower state energy errors\label{tab:errors}}
\tablehead{ \multirow{2}{*}{Line\tablenotemark{a}} & \colhead{Wavenumber} & \colhead{Calculated $E_{low}$\tablenotemark{b}} & \colhead{HITRAN $E_{low}$} & \colhead{Difference} \\ & \colhead{(cm$^{-1}$)} & \colhead{(cm$^{-1}$)} & \colhead{(cm$^{-1}$)} & \colhead{\%} }
\startdata
$\tilde\nu[1]$ & 1313.30 & 1364.00 & 1336.88 & 2.0 \\
$\tilde\nu[2]$ & 3072.57 & 1590.79 & 1494.72 & 6.4 \\
$\tilde\nu[3]$ & 1323.51 & 1644.41 & 1639.18 & 0.3 \\
$\tilde\nu[4]$ & 1304.20 & 1879.95 & 1780.71 & 5.6 \\
$\tilde\nu[5]$ & 1301.59 & 2036.83 & 1976.72 & 3.0 \\
\enddata
\tablenotetext{a}{These lines are displayed in Figure~\ref{fig:slope}.}
\tablenotetext{b}{As shown in the slope of the lines in Figure~\ref{fig:slope}.}
\end{deluxetable}

\begin{deluxetable}{ccccc}
\tablenum{8}
\tabletypesize{\small}
\tablewidth{0pt}
\tablecaption{Extract of the 1000$^{\circ}$C hot CH$_{4}$ line list\tablenotemark{a}\label{tab:extract}}
\tablehead{ \colhead{Temperature}  &  \colhead{Wavenumber} &   \colhead{Intensity}   &  \colhead{$E_{\textrm{low}}$}  & \colhead{Quality} \\
           \colhead{($^{\circ}$C)} & \colhead{(cm$^{-1}$)} & \colhead{(cm molecule$^{-1}$)} & \colhead{(cm$^{-1}$)} & \colhead{Factor} }
\startdata
 ...&         ...&      ...&      ...& ...\\
1000& 1343.952393& 4.21E-22& 3.29E+03&   1\\
1000& 1343.962722& 1.31E-23& 1.82E+03&   H\\
1000& 1343.974972& 1.98E-23& 2.17E+03&   H\\
1000& 1343.980835& 8.99E-23& 3.08E+03&   2\\
1000& 1343.982910& 8.99E-23& 2.78E+03&   1\\
1000& 1343.992990& 2.85E-23& 1.93E+03&   H\\
1000& 1344.015503& 3.96E-23& 5.20E+03&   3\\
1000& 1344.033569& 1.19E-22& 3.18E+03&   1\\
 ...&         ...&      ...&      ...& ...\\
\enddata
\tablenotetext{a}{ The complete line list is available along with the erratum (2013, \textit{ApJ}, 774, 89, doi:10.1088/0004-637X/774/1/89)}
\end{deluxetable}

\section{DISCUSSION}

We have produced CH$_{4}$ line lists which can be used at temperatures from 300$^{\circ}$C to 1400$^{\circ}$C. The calibration of the wavenumber scale for each spectrum was done by making wavenumber comparisons of 20 lines to the HITRAN database and this gives an overall accuracy of better than 0.005 cm$^{-1}$. The wavenumber calibration is straightforward but intensity calibration is more difficult. Line intensities are notoriously hard to calibrate and our calibration was made by comparing as many lines as possible to the HITRAN database. Our overall line intensity accuracy is estimated to be approximately a factor of two. Intensity matches for lower wavenumbers (regions A and B) were good and allowed a consistent calibration function to be determined. Each region overlapped the neighbouring region with enough emission lines to place all regions on the same scale. For region D the intensity matches with HITRAN were effectively absent so a calibration factor could not be determined from this region alone. Since all regions are on the same scale, the calibration function was extrapolated into region D. In order to validate our extrapolation we were able to compare our 1000$^{\circ}$C spectrum with the equivalent 1273 K spectrum from \citet{nassar03}. The matches ($\pm$0.0025 cm$^{-1}$) can be seen in Figure~\ref{fig:nassar} and show a clear intensity correlation to within a factor of two which gives confidence in the calibration applied. The slope is present because the \citet{nassar03} calibrations were performed using a single factor whereas the calibration used here is a parabolic function.

\begin{figure}[t]
\epsscale{1}
\plotone{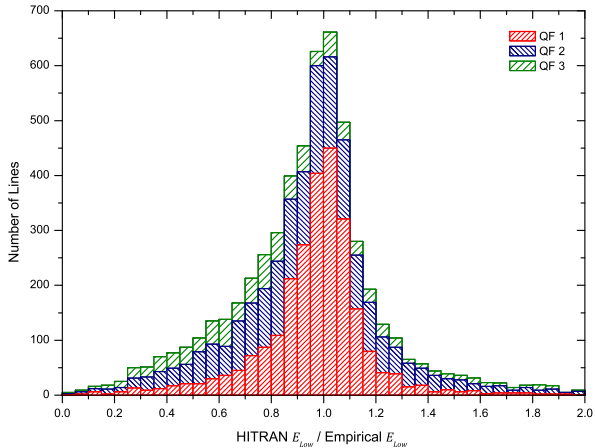}
\caption{Histogram showing the ratios of the HITRAN $E_{low}$ values to the empirical $E_{low}$ values. The vertical bars of the histogram have been shaded to indicate the number of points with the quality factors (QF) 1, 2 and 3.\label{fig:elowerrors}}
\end{figure}

It is difficult to estimate an error for the empirical $E_{low}$ values presented here because self absorption causes inaccurate intensities resulting in inaccurate empirical $E_{low}$ values. This means comparisons can only be made between emission lines removed from the line lists and lines inserted from the HITRAN database. Comparison of the $E_{low}$ values of the five lines in Figure~\ref{fig:slope} can give a false impression of the accuracy, as in Table~\ref{tab:errors}. A better estimate can be obtained by plotting the ratio of $E_{low}$ from HITRAN compared to the empirical $E_{low}$ values of the removed lines (Figure~\ref{fig:elowerrors} at 1000$^{\circ}$C). It clearly shows that the most common value of the HITRAN $E_{low}$ /empirical $E_{low}$ ratio is approximately one and that the majority of empirical $E_{low}$ values are accurate to within 20 \%. It can be seen that the most accurate lines (with a quality factor of `1') also display the smallest spread as opposed to the least accurate lines (quality factor of `3').  Figure~\ref{fig:elowerrors} also shows why it was necessary to remove these lines because there is a tendency towards low ratios ($<1$) due to inaccurate line intensities caused by self absorption.

At high temperatures, thermal decomposition of methane is expected \citep{thievin08} and lower intensities are indeed observed for our high temperature spectra. The highest temperature spectra recorded in region D (1200$^{\circ}$C and 1300$^{\circ}$C) were contaminated by the absorption and emission of the products of thermal decomposition, namely C$_{2}$H$_{2}$ and C$_{2}$H$_{4}$. Attempts have been made to remove these species by using recent line lists \citep{moudens11, rothman09}; however, the emission features could not be completely removed. As this is the case, the hot CH$_{4}$ emission lines could not be distinguished from the products of thermal decomposition and the 1200$^{\circ}$C and 1300$^{\circ}$C files of region D have been excluded from our analysis.

Empirical $E_{low}$ values have been determined previously using a similar method by our group for the hot line lists of NH$_{3}$ \citep{hargreaves11, hargreaves12}. It has also been used in the submillimeter regime \citep{fortman10} and on `cold' CH$_{4}$ in the infrared \citep{sciamma09,lyulin10,wang12}. \citet{fortman10} obtained $E_{low}$ values of an astrophysical `weed' (ethyl cyanide) using numerous cool (-$40$ -- $140^{\circ}$C) spectra. \citet{sciamma09}, \citet{lyulin10} and \citet{wang12} have used a similar approach to study CH$_{4}$ based on a small number of room temperature and cold ($\sim$80 K) spectra. The method presented here uses a dozen `hot' (300 -- 1400$^{\circ}$C) spectra which were recorded under conditions that are similar to those found in the atmospheres of brown dwarfs and hot-Jupiters. The use of a large number of spectra improves the accuracy of the $E_{low}$ values. However a single spectrum takes us many hours to record (including heating and cooling of the system) so it is impractical to record many additional spectra. Nevertheless, combining both hot and cold CH$_{4}$ line lists would help to complete the knowledge of CH$_{4}$ in the infrared.

Our method produces a relatively small number of lines with high line position accuracy. An alternative approach is to use \textit{ab initio} calculations which provide significantly more lines but the line positions have much lower accuracy. The two methods are complementary and they can be combined to produce a merged line list. \citet{zobov11} have used the \textit{ab initio} approach and were able to assign quantum numbers to some of our new NH$_{3}$ data; similar calculations are currently underway for CH$_{4}$.\\

The CH$_{4}$ line lists are available as electronic supplements via the website of the Astrophysical Journal (2013, \textit{ApJ}, 774, 89, doi:10.1088/0004-637X/774/1/89) and spectra can also be provided upon request. The line lists should be used at the specified temperature; if a temperature is needed between those provided, we recommend scaling the lines using Equation~\ref{eqn:ratios}. Only line lists from this erratum should be used.

\section{CONCLUSION}

Experimental line lists of hot CH$_{4}$ have been provided which contain calibrated wavenumbers (in cm$^{-1}$), calibrated intensities (in cm molecule$^{-1}$) and empirical lower state energies (in cm$^{-1}$). The line lists were based on 12 emission spectra recorded at 300 -- 1400$^{\circ}$C in 100$^{\circ}$C intervals using a FT-IR spectrometer and cover the 960 -- 5000 cm$^{-1}$ wavenumber range. This range contains the dyad, pentad and octad infrared regions of CH$_{4}$. The line lists can be used directly to model CH$_{4}$ in brown dwarfs and exoplanets. They also provide new experimental data for comparison with theoretical predictions.

\acknowledgments

Support for this work was provided by a Research Project Grant from the Leverhulme Trust and a Department of Chemistry (University of York) studentship. LM was supported by a Nuffield Foundation Undergraduate Research Bursary.\\

The authors would like to thank Michael Rey (Universit\'{e} de Reims) who brought the problem of the original data to our attention.

\clearpage

\end{document}